\documentclass[preprint,amsmath,amssymb,aps,superscriptaddress,nofootinbib,prd]{revtex4-1}
\usepackage{graphicx}
\usepackage{color}

\begin{document}


\title{Linking the Gauge Hierarchy with Neutrino Masses\\ and Dark Matter via Two-step Cosmological Selection}

\author{Jin-Lei Yang}\email{jlyang@hbu.edu.cn}
\affiliation{Department of Physics, Hebei University, Baoding, 071002, China}
\affiliation{Key Laboratory of High-precision Computation and Application of Quantum Field Theory of Hebei Province, Baoding, 071002, China}
\affiliation{Research Center for Computational Physics of Hebei Province, Baoding, 071002, China}

\author{Frank F. Deppisch}\email{f.deppisch@ucl.ac.uk}
\affiliation{Department of Physics and Astronomy, University College London, London WC1E 6BT, United Kingdom}

\date{\today}

\begin{abstract}
The hierarchy problem between the electroweak (EW) and Planck scales remains a central puzzle in modern physics. We discuss a promising solution operating through the cosmological selection of the EW vacuum in a multiverse landscape, where the EW scale is dynamically approached as the configuration that maximizes the vacuum energy. By extending the Standard Model with a complex scalar singlet and right-handed neutrinos, charged under a global $U(1)_{B-L}$ symmetry, the model not only explains the smallness of the EW scale. It can also account for neutrino masses via the seesaw mechanism and the matter–antimatter asymmetry via leptogenesis. In addition, it provides a viable dark matter candidate that is testable in future neutrino experiments.
\end{abstract}

\maketitle


\section{Introduction}
\label{sec:intro}

The spontaneous breaking of electroweak (EW) symmetry, triggered by a non-zero vacuum expectation value (VEV) of the Higgs field, provides masses to the elementary particles in the Standard Model (SM). The SM Higgs potential can be written as
\begin{align}
    V_H = 
    - \frac{1}{2}\mu_H^2 H^\dagger H 
    + \frac{1}{4}\lambda_H (H^\dagger H)^2,
\label{eq1}
\end{align}
where $\mu_H^2 > 0$, $\lambda_H > 0$ and the VEV is $\langle H\rangle = v \approx 246$~GeV. The discovery of the 125~GeV Higgs boson at the Large Hadron Collider (LHC)~\cite{ATLAS:2012yve, CMS:2012qbp} confirms the remarkable success of this mechanism. However, the origin of the EW VEV being so many orders of magnitude smaller than the Planck scale remains unexplained, thereby giving rise to the well-known gauge hierarchy problem~\cite{Gildener:1976ai, Weinberg:1978ym}. The gauge hierarchy problem is usually formulated such that the Higgs mass is unprotected against large quantum corrections from high energy scales, rendering its small value technically unnatural.

To solve the hierarchy problem, various approaches have been considered. Supersymmetry posits that every fundamental particle has a superpartner with a different spin~\cite{Golfand:1971iw, Ramond:1971gb, Wess:1973kz, Wess:1974tw}. This ensures that the quantum corrections to the Higgs mass from fermions and bosons cancel out precisely, unless supersymmetry-breaking contributions are included. Theories with extra dimensions suggest that the fundamental Planck scale is much closer to the EW scale than it appears~\cite{Arkani-Hamed:1998jmv}. Composite Higgs models theorize that the Higgs boson is not a fundamental particle, but is composed of constituents~\cite{Contino:2003ve, Agashe:2004rs}. 

Apart from the dynamics of specific particle physics theories, the anthropic principle suggests that our universe is part of a vast multiverse and we simply happen to inhabit the statistically unlikely universe where the parameters are tuned for a light Higgs mass, because only this type of universe can support the complex structures needed for observers to exist~\cite{Agrawal:1997gf}. Using the multiverse landscape, cosmological relaxation~\cite{Graham:2015cka} introduces a dynamic scalar field, the relaxion, that rolled down a potential energy gradient in the early universe, naturally settling the Higgs mass to its current observed value without the need for fine-tuning. Another example is the crunching dilaton model~\cite{Csaki:2020zqz}, which dynamically selects the EW scale by tying the Higgs VEV to cosmic stability where only small Higgs VEVs generate a metastable vacuum avoiding a cosmic crunch, enabling viable cosmic expansion. Other ideas that solve the hierarchy problem within the multiverse landscape can be found in Refs.~\cite{Giudice:2019iwl, Strumia:2020bdy, TitoDAgnolo:2021nhd, TitoDAgnolo:2021pjo, Matsedonskyi:2023tca, Benevedes:2025qwt}.

Based on the observation that the regions with maximal vacuum energy will eventually dominate the multiverse in volume~\cite{Geller:2018xvz}, a cosmological solution to the gauge hierarchy problem is proposed in a model with two Higgs doublets and a pseudoscalar singlet~\cite{Chattopadhyay:2024rha}. In our analysis, we find that the same goal can be achieved by a two-step cosmological selection mechanism, with a much simpler setup, namely the complex scalar singlet $\Phi$ extended SM with a global $U(1)_{B-L}$ symmetry. More notably, its implications extend far beyond the gauge hierarchy problem. It incorporates a DM candidate and by naturally incorporating three generations of right-handed neutrinos, it can potentially address the issue of neutrino mass generation and the matter-antimatter asymmetry.

Our proposed two-step cosmological selection mechanism relates closely to the history of the universe. When temperature is extremely high, $T \gg v_\phi$, both the SM and $U(1)_{B-L}$ symmetries remain unbroken. As the universe cools down, $T \approx v_\phi$, $\Phi$ acquires a large VEV at the first step. Approaching the EW scale, $T \approx 100$~GeV, the universe evolves into a multiverse landscape of SM VEVs, $\langle H\rangle = v_1, v_2, v_3, ...\neq 0$. The proposed mechanism naturally predicts the vacuum energy is maximized for a small but non-zero SM Higgs VEV, and the corresponding bubble will eventually dominate the multiverse.

Since right-handed neutrinos (RHNs) are included naturally with respect to the $U(1)_{B-L}$ symmetry, the remarkably economical model provides elegant solutions to four outstanding puzzles in particle physics : (1) the gauge hierarchy problem is solved by the proposed  two-step cosmological selection mechanism; (2) the observed neutrino oscillation is explained via type-I seesaw mechanism~\cite{Minkowski:1977sc,Weinberg:1979sa}; (3) the matter-antimatter asymmetry arises from leptogenesis~\cite{Fukugita:1986hr}; (4) the CP-odd component of $\Phi$ serves as a viable dark matter (DM) candidate. Most importantly, the predicted DM candidate is testable at detectors such as JUNO~\cite{Akita:2022lit}, DUNE~\cite{Arguelles:2019ouk}, and HyperKamiokande~\cite{Bell:2020rkw} by the process DM decays to active neutrinos.

The paper is organized as follows. In Sec.~\ref{sec:model}, we define the model accounting for the puzzles mentioned above. Sec.~\ref{sec:neutrino} gives the neutrinos sector of the model. The proposed cosmological selection mechanism is presented in Sec.~\ref{sec:TCS}. The relic abundance of viable DM candidate in the model is calculated in Sec.~\ref{sec:dm}. The discussions about the predicted numerical results and conclusions are made in Sec.~\ref{sec:conclusions}.


\section{Model}
\label{sec:model}
The proposed model extends the SM by an additional global $U(1)_{B-L}$ symmetry, a complex scalar singlet $\Phi$ and three RHNs $\nu_{R_i}$. The $U(1)_{B-L}$ charges of $\Phi$ and $\nu_{R_i}$ are $2$ and $-1$, respectively. While $\Phi$, $\nu_{R_i}$ do not carry any SM gauge charges, the other SM fields have their canonical $B-L$ charges. The Lagrangian of the model is given by
\begin{align}
    \mathcal{L} 
    &= \mathcal{L}_\text{SM} \nonumber\\
    &+ \frac{1}{2}\mu_\Phi^2\Phi\Phi^* 
     - \frac{1}{4}\lambda_\Phi (\Phi\Phi^*)^2
     - \lambda_{H\Phi} (\Phi\Phi^*) (H^\dagger H)
     - V_\text{soft} \nonumber\\
    &+ Y^R_{ij}\Phi \bar\nu_{R_i}^c \nu_{R_j} 
     + Y^D_{\alpha j} \bar L_\alpha \cdot H \nu_{R_j},
\label{eq2}
\end{align}
where $\mathcal{L}_\text{SM}$ is the SM Lagrangian, including the Higgs potential in Eq.~\eqref{eq1}. The scalar potential is extended by the terms on the second line with $\mu_\Phi^2 \gg \mu_H^2$, $\lambda_{H\Phi}$ is a real coupling constant and $\lambda_\Phi > 0$. The contribution $V_\text{soft}$ includes terms that softly break the $U(1)_{B-L}$ symmetry which we will discuss later. The third line contains the additional Yukawa couplings of the EW lepton doublets $L_\alpha = (\nu_L, e_L)_\alpha^T$  ($\alpha = e,\mu,\tau$) and the RHNs.

The scalar potential without $V_\text{soft}$ defined in Eq.~\eqref{eq2} corresponds precisely to the complex singlet scalar extension of the SM (CxSM)~\cite{Barger:2008jx} with a global $U(1)$ symmetry. One of the primary motivations for the CxSM is to accommodate a DM candidate. Upon spontaneous breaking of the $U(1)$ symmetry, a massless Nambu-Goldstone boson emerges and as such would not be a DM candidate. Therefore, soft terms explicitly breaking the $U(1)$ symmetry must be introduced to endow the would-be Goldstone boson with a nonzero mass, rendering it a viable cold DM candidate~\cite{Gross:2017dan, Chiang:2017nmu, Cheng:2018ajh, Chen:2019ebq, Cho:2021itv, Cho:2022zfg, Schicho:2022wty, Zhang:2023mnu, Pham:2024vso, Lane:2024vur}. The most general form of $V_\text{soft}$ reads
\begin{align}
    -V_\text{soft} 
    =& \kappa_1^3\Phi 
     + \kappa_2^2 \Phi^2
     + \kappa_3 \Phi^3 
     + \kappa_4\Phi^2\Phi^*\nonumber\\
    &+\kappa\Phi (H^\dagger H) + \text{h.c.},
\end{align}
where $\kappa$, $\kappa_i$, $(i=1,..,4)$ are coupling constants with units of energy. DM candidate then receives a tree-level nonzero mass through the terms in $V_\text{soft}$. Generally, $\kappa$ and $\kappa_i$ can be complex, potentially affecting the stability of the DM candidate $A_\phi$ $(\Phi\equiv v_\phi + S_\phi + i A_\phi)$. To obtain a viable DM candidate, mixing between the SM-like Higgs and $A_\phi$ must be avoided. This is achieved by taking all $\kappa_i$ to be real \cite{Barger:2008jx} which is our choice in the following. In this case, the Hermiticity of the potential imposes a $Z_2$ symmetry for the CP-odd component $A_\phi$ of $\Phi$, which makes $A_\phi$ stable.

To gain an intuition for the size of the parameters in this model, we consider a benchmark point as
\begin{gather}
    v_\phi = 10^{12}~\text{GeV},\;
    \lambda_\Phi = 0.1,\;
    \lambda_{H\Phi} = -3.9\times10^{-21}, \nonumber\\
    \kappa, \kappa_i = 5\times10^{-14}~\text{GeV},
\label{eqBM}
\end{gather}
For the benchmark point defined above, we assume $\kappa = \kappa_i$. This simplifying assumption does not affect the physical predictions of the model. The negative value of $\lambda_{H\Phi}$ is required and calculated in our cosmological selection mechanism as detailed below. We set $\lambda_\Phi = 0.1$, similar to the SM coupling $\lambda_H$. According to the t'Hooft naturalness principle~\cite{tHooft:1979rat}, parameters can be naturally small if setting them to zero increases the symmetry of the theory. Hence, $\kappa$ and $\kappa_i$ are naturally small. For the portal coupling $\lambda_{H\Phi}$, its $\beta$ function is proportional to itself or neutrino couplings. At one-loop, the renormalization group evolution has the structure $16\pi^2 \partial\lambda_{H\Phi} / (\partial \ln\mu) = \lambda_{H\Phi}(3\lambda_H - \frac{9}{10}g_1^2 + 2\sum_{\alpha}|Y_{e_\alpha}|^2 + \dots) - 16\,\text{Tr}(Y_D\cdot Y_R^* \cdot Y_R \cdot Y_D^\dagger)$, where $g_1$ denotes the $U(1)_Y$ gauge coupling and $Y_{e_\alpha}$ are the charged lepton Yukawa couplings ($\alpha = e, \mu, \tau$) shown as a typical contribution. The contribution from neutrino couplings can always be neglected safely when RHNs are lighter than $\approx 10^7~\text{GeV}$.


\section{Neutrino Sector}
\label{sec:neutrino}
The newly introduced Yukawa couplings in Eq.~\eqref{eq2} indicate $\Phi$ plays the role of a singlet Majoron~\cite{Chikashige:1980ui}, where $v_\phi$ serves as the origin of Majorana masses for the RHNs after the spontaneous breaking of $U(1)_{B-L}$ symmetry. The observed small neutrino masses are then naturally explained via the type-I seesaw mechanism~\cite{Minkowski:1977sc,Weinberg:1979sa},
\begin{align}
    m_\nu \approx -M_D^T \cdot M_R^{-1} \cdot M_D,
\end{align}
with $M_D = Y_D v$ and $M_R = Y_R v_\phi$. Fitting the neutrino masses predicted by the type-I seesaw mechanism to the experimental observations~\cite{Casas:2001sr, Ibarra:2003up, Gu:2006wj, Hirsch:2009ra, CarcamoHernandez:2019eme, Xing:2020ald, Yang:2024duo, Chen:2025cor}, and explaining the observed matter-antimatter asymmetry in the universe through leptogenesis~\cite{Davidson:2002qv, Giudice:2003jh, Pilaftsis:2003gt, Davidson:2008bu, Canetti:2012kh, Drewes:2016jae, Klaric:2020phc, Bhalla-Ladd:2025agq, Takada:2025epa}, have been widely studied. It is worth mentioning that, for the case of non-degenerate RHN masses, successful thermal leptogenesis requires the lightest RHN mass $M_{N_1} \gtrsim 6.7 \times 10^{8}$~GeV with three RHN generations~\cite{Okada:2025daq}. Perturbativity of the corresponding Yukawa couplings implies $v_\phi \gtrsim M_{N_i}$ and the lower bound on $v_\phi$ can be relaxed if two or more RHNs are degenerate~\cite{Iso:2010mv, Dev:2017wwc}.


%
\begin{figure*}[t!]
    \includegraphics[width=0.49\textwidth]{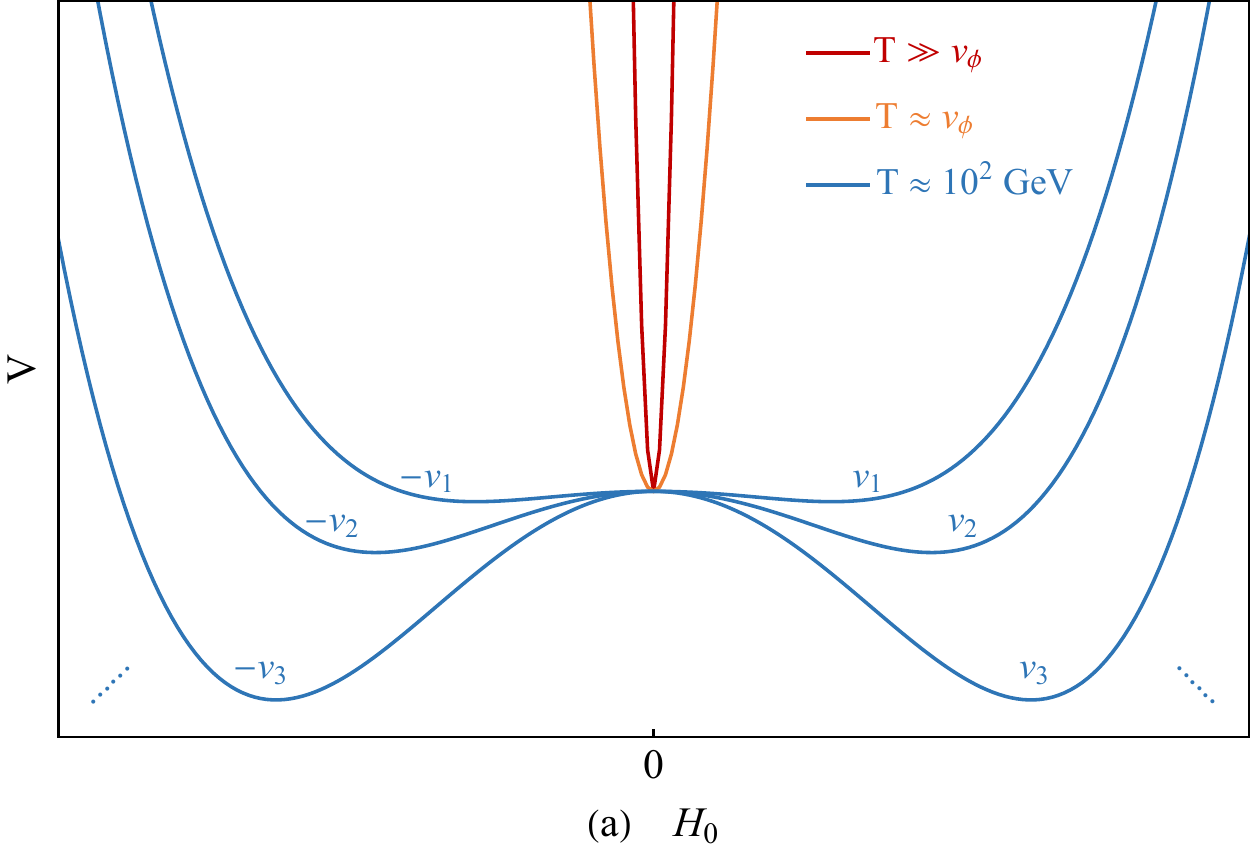}
    \includegraphics[width=0.49\textwidth]{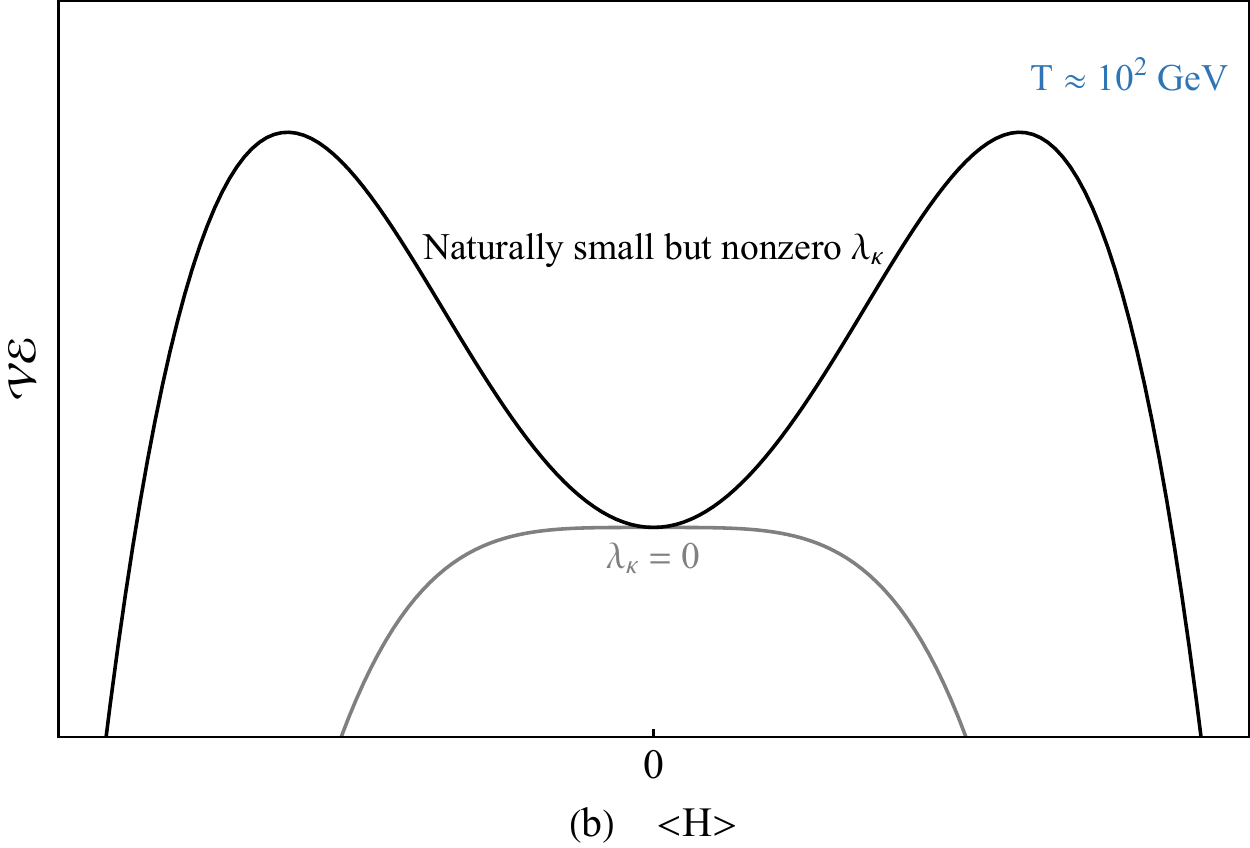}
    \caption{Sketch of the scalar potential $V(H)$ of the SM Higgs field at different temperatures (a) and the total vacuum energy $\mathcal{V}\mathcal{E}$ as a function of the SM Higgs VEV $\langle H\rangle$, when cosmological selection occurs at the second step.}
\label{Fig1}
\end{figure*}

\section{Cosmological Selection Mechanism}
\label{sec:TCS}
For positive $\mu_\Phi^2$, the scalar potential defined in Eq.~\eqref{eq2} takes the minimum at $\langle H\rangle = \langle \Phi\rangle = 0$ when the temperature is high ($T\gg v_\phi$), shown by the dark red curve in Fig.~\ref{Fig1}~(a). This is because the leading-order thermal corrections to the squared mass terms $-\mu_H^2$ and $-\mu_\Phi^2$ are of the order $\mathcal{G}_H T^2$, $\mathcal{G}_\Phi T^2$, respectively. Here, $T$ denotes temperature, while $\mathcal{G}_H$, $\mathcal{G}_\Phi$ are thermal coupling coefficients that capture the relevant interactions of $H$ and $\phi$~\cite{Parwani:1991gq, Carrington:1991hz, Quiros:1999jp, Hashino:2018zsi}. Given that the couplings of $\Phi$ are weaker than the ones of $H$, we have that $\mathcal{G}_\Phi \ll \mathcal{G}_H$. This hierarchy, combined with $\mu_\Phi^2 \gg \mu_H^2$, ensures that as the universe cools ($T\approx v_\phi$), it first undergoes a transition into a phase with $\langle H\rangle = 0$, $\langle \Phi\rangle \neq 0$, indicated by the dark orange curve in Fig.~\ref{Fig1}~(a). This concludes the first step of the cosmological selection mechanism.

As the temperature continues to drop ($T\approx 100$~GeV), the SM Higgs doublet acquires its nonzero VEV and the SM gauge group breaks down to the $U(1)_\text{EM}$ of electromagnetic symmetry. The universe evolves into a landscape of vacua characterized by a given $v_\phi$ but a continuum of SM Higgs VEVs $\langle H\rangle = v_1, v_2, v_3, ...$, indicated by the blue curves in Fig.~\ref{Fig1}~(a). After applying the tadpole equations obtained by minimizing the full scalar potential, the multiverse will be dominated in volume by the bubble with maximal vacuum energy. The total vacuum energy can be expressed in terms of the SM Higgs VEV $v_*$ in a given vacuum domain as
\begin{align}
    {\mathcal V}{\mathcal E} 
    = &- \frac{1}{4}\lambda_H v_*^4 
       + \lambda_\kappa v_\phi^2 v_*^2 
       - \frac{1}{4}\lambda_\phi v_\phi^4 \nonumber\\
      &- \kappa_1^3 v_\phi 
       + (\kappa_3 
       + \kappa_4)v_\phi^3,
\label{eqa13}
\end{align}
with
\begin{align}
    \lambda_\kappa &= 
        \frac{\kappa}{v_\phi}-\lambda_{H\Phi}, \nonumber\\
    \mu_H^{2} &= 
        \lambda_H v_*^2 + 2v_\phi\left(\lambda_{H\Phi} v_\phi - 2\kappa\right), 
        \nonumber\\
    \mu_\Phi^{2} &= 
        2\lambda_{H\phi}v_*^2 + \lambda_\Phi v_\phi^2 - v^2\frac{2\kappa}{v_\phi}
        - \frac{2\kappa_1^3}{v_\phi} \nonumber\\
        &-4\kappa_2^2-6\left(\kappa_3 + \kappa_4\right)v_\phi. \label{eqa16}
\end{align}
Maximizing $\mathcal{V}\mathcal{E}$ with respect to $v_*$ yields
\begin{align}
    v_*^\text{max} = \sqrt{\frac{2\lambda_\kappa}{\lambda_H}}v_\phi,
\label{eqa19}
\end{align}
Among the various $v_\star = v_1, v_2, v_3,\dots$, if the specific configuration $\langle H\rangle = v_*^\text{max}$ maximizes the vacuum energy as shown by the black curve in Fig.~\ref{Fig1}~(b), the vacuum domains (bubbles) with $\langle H\rangle =  v_*^\text{max}$ will eventually dominate the volume of the multiverse. This completes the final step of the proposed mechanism. 

Therefore, the key to this mechanism is that $\langle H\rangle = v$ maximizes the vacuum energy at the second step. But when $\lambda_\kappa = 0$, the case refers to a typical vacuum energy decreases with increasing $\langle H\rangle$, as shown by the gray curve in Fig.~\ref{Fig1}~(b). For example, the vacuum energy of the SM scalar potential is proportional to $-\lambda_H \langle H\rangle^4$ after applying the tadpole equation, thus maximizing the vacuum energy favors small $\langle H\rangle$. This fact provides a cosmological solution to the hierarchy problem~\cite{Chattopadhyay:2024rha}. However, this reasoning alone cannot determine the magnitude of $\langle H\rangle$, because the vacuum energy increases as $\langle H\rangle \to 0$. Consequently, the global maximum would naively be at $\langle H\rangle = 0$, not at the EW scale $v$.

For the mechanism proposed in this work, a nonzero coupling $\lambda_\kappa$ defined in Eq.~\eqref{eqa16} plays the role of preventing the trivial vacuum $\langle H \rangle \to 0$ from being the global energy maximum. Once $\lambda_{H\Phi}$ is chosen to be small at a given scale, it remains small under renormalization group running, protected from receiving large corrections. Hence, $\lambda_\kappa$ is naturally small and the vacuum energy takes its maximum at a small SM Higgs VEV when $\lambda_\kappa > 0$, as shown by the black curve in Fig.~\ref{Fig1}~(b). As a result, a small ($v \ll v_\Phi$) but nonzero EW scale is predicted naturally by the two-step cosmological selection mechanism. 

Imposing the observed EW scale $v_* = v\approx 246$~GeV in Eq.~\eqref{eqa19} to Eq.~\eqref{eqa16}, we obtain
\begin{align}
    \lambda_{H\Phi} &=\frac{\kappa}{v_\phi}-\frac{\lambda_H v^2}{2v_\phi^2} \nonumber\\
    \mu_H^{2} &= -2\kappa v_\phi, \nonumber\\
    \mu_\phi^2 &= \lambda_\Phi v_\phi^2 
    - \lambda_{H}v^4/v_\phi^2 - 4\kappa_2^2 
    - 6\left(\kappa_3+\kappa_4\right)v_\phi^2. \label{eqaa22}
\end{align}
It can be noted that when $\kappa = 0$, the portal coupling $\lambda_{H\Phi}$ naturally generates the negative squared mass term for the SM Higgs doublet. This result is analogous to the one obtained in the classically scale-invariant theory with a Higgs portal~\cite{Coleman:1973jx, Binoth:1996au, Schabinger:2005ei, Patt:2006fw, Englert:2011yb, Englert:2013gz, Khoze:2013oga}, where the EW symmetry breaking scale is induced radiatively by the Coleman-Weinberg mechanism, and the smallness of the Higgs VEV is explained naturally through the added Higgs portal. In the present framework, this outcome is not assumed by the scale symmetry, but follows directly from the proposed mechanism in the limit $\kappa = 0$.


\section{Dark Matter}
\label{sec:dm}
Applying Eq.~\eqref{eqaa22}, we can obtain the physical scalar mass squared eigenvalues approximately as
\begin{align}
    M_h^2 &\approx 2\lambda_H v^2\approx(125~\text{GeV})^2, \nonumber\\
    M_\phi^2 &\approx 2\lambda_\Phi v_\phi^2 
                      + \frac{2\kappa v^2}{v_\phi}
                      + \frac{2\kappa_1^3}{v_\phi} 
                      - 6\left(\kappa_3+\kappa_4\right) v_\phi, \nonumber\\
    M_A^2 &= \frac{2\kappa v^2}{v_\phi} + \frac{2\kappa_1^3}{v_\phi}
            + 8\kappa_2^2 + \left(18\kappa_3 + 2\kappa_4\right)v_\phi, \label{eq15}
\end{align}
where $v_\phi \gg v$ is assumed to obtain $M_h^2$ and $M_\phi^2$. For comparable magnitudes of $\kappa$ and $\kappa_i$, the mass of the new CP-even scalar is dominated by the term $M_\phi^2 \approx 2\lambda_\phi v_\phi^2$, whereas the mass of the CP-odd scalar scales as $M_A^2 \approx (18\kappa_3 + 2\kappa_4)v_\phi$. Consequently, nonzero $\kappa_3$ and $\kappa_4$ significantly affect the DM mass $M_A$. The corresponding mass eigenstates $h$ (SM-like Higgs), $A$ (CP-odd scalar), and $\phi$ (heavy CP-even scalar) are related to the gauge eigenstates $S_H$, $S_\phi$, and $A_\phi$ by
\begin{gather}
    h \approx S_H + \frac{\lambda_H v^3}{\lambda_\phi v_\Phi^3}S_\phi, \nonumber\\
    \phi \approx -\frac{\lambda_H v^3}{\lambda_\Phi v_\phi^3}S_H + S_\phi,\qquad 
    A = A_\phi,
\label{eq16}
\end{gather}
with $S_H$ denoting the neutral CP-even component of the SM Higgs doublet $H$.

For the benchmark in Eq.~\eqref{eqBM}, the DM candidate $A$ does not equilibrate with the SM through the portal coupling $\lambda_{H\Phi}$~\cite{Bernal:2017kxu}, and would typically be classified as feebly interacting massive particle (FIMP) DM. Hence, the predicted mass and coupling pattern aligns well with the conditions required for ultra-relativistic freeze-out (UFO)~\cite{Henrich:2025sli, Henrich:2025gsd, Henrich:2025pca}, a production mechanism wherein DM attains its relic abundance during the reheating epoch. Relativistic freeze-out during the radiation dominated era by assuming instantaneous reheating is also valid (in a similar manner to neutrinos), but the DM mass is required to be very light to saturate the observed relic abundance~\cite{Henrich:2025gsd}, and the resulting relic density is essentially independent of the reheating temperature. Moreover, this kind of DM candidate as weakly interacting massive particle is widely studied in the CxSM~\cite{Gross:2017dan, Chiang:2017nmu, Cheng:2018ajh, Chen:2019ebq, Cho:2021itv, Cho:2022zfg, Schicho:2022wty, Zhang:2023mnu, Pham:2024vso, Lane:2024vur}. The main focus of this work is therefore on the UFO during the reheating epoch, where the interplay between DM production and the dynamics of reheating leads to interesting phenomenological consequences.

The dominant DM production arises from the processes $\phi\phi \to AA$ and $\phi \to AA$.\footnote{When DM is produced in UFO, the contributions from $\nu_R\nu_R \to AA$ to the production rate are negligible compared to the ones from $\phi\phi \to AA$ and $\phi \to AA$.} The calculation of the relevant production rate $R_A$ can be found in the appendix, with the rate dominantly driven by the quantity $\lambda_\Phi v_\phi$. Hence, for the calculation of the DM relic density, the main impact of the $\kappa_i$ is mainly on the DM mass. The evolution of the DM number density $n_A$ is governed by the relevant Boltzmann equation, conveniently expressed in terms of the co-moving yield $Y_A \equiv n_A a^3$ as~\cite{Cosme:2021baj}
\begin{align}
    \frac{dY_A}{da} = 
    \frac{2a^2}{H(a)}\left(1-\frac{Y_A^2}{(Y_A^{\rm eq})^2}\right)R_A.
\label{eq25}
\end{align}
Here, $a$ is the cosmological scale factor, and $Y_A^\text{eq} = n_A^\text{eq} a^3$ is the yield in the thermal equilibrium. During the reheating era, for simplicity, the Hubble rate $H(a)$ is taken as~\cite{Kaneta:2019zgw}
\begin{align}
    H(a) = 
    \sqrt{\frac{427\pi^2}{360}} 
    \frac{T_\text{RH}^2}{M_\text{Planck}}\left(\frac{a_\text{RH}}{a}\right)^{3/2}
\end{align}
where $a_\text{RH}/a = (T/T_\text{RH})^{8/3}$ was applied, the reduced Plank mass is $M_P = 2.4\times 10^{18}$~GeV and $T_\text{RH}$ denotes the reheating temperature with lower bound $T_\text{RH} > 4$~MeV from Big Bang Nucleosynthesis constraints~\cite{Hannestad:2004px}. Solving Eq.~\eqref{eq25} gives the co-moving yield at the end of reheating, $Y_A(a_\text{RH})$. By requiring the present-day DM relic abundance $\Omega_Ah^2 = 0.12$, the DM mass is approximately related to the breaking scale $v_\phi$ and $T_\text{RH}$ from solving the Boltzmann evolution as
\begin{align}
    M_A \approx 
    \left(\frac{T_\text{RH} / v_\phi}{10^{-3}}\right)^{-6}
    \times\text{GeV},
\end{align}
c.f. Fig.~\ref{fig3}~(a) discussed below. As can be seen in Eq.~\eqref{eqBM}, the couplings between $A$ and the SM sector are extremely weak. Consequently, $A$ readily evades constraints from direct (e.g., scattering processes of DM on nuclei), indirect (e.g., astrophysical DM annihilation) and collider (e.g., missing energy signatures) searches. On the other hand, $A$ is not stable as it decays, dominantly to light neutrinos at tree-level via the active-sterile mixing with the RHN, $A\to\nu\nu$. The DM life-time is given by~\cite{Mohapatra:2021ozu}
\begin{align}
    \tau_A &= 
    \left[\frac{M_A}{4\pi v_\phi^2} 
    \sum_{\alpha,\beta=1}^3
    \left|\left(M_D^T \cdot  M_R^{-1} \cdot M_D\right)_{\alpha\beta}\right|^2\right]^{-1} 
    \nonumber\\
    &= \frac{1~\text{GeV}}{M_A}
    \frac{10^{-2}~\text{eV}^2}{\sum_{i=1}^3 m_{\nu_i}^2}
    \left(\frac{v_\phi}{10^{12}~\text{GeV}}\right)^2 \times 10^{23}~\text{sec},
\end{align}
with the light neutrino masses $m_{\nu_i}$. This directly connects the DM phenomenology with neutrino observables. Using neutrino oscillation data \cite{Esteban:2024eli}, we have $\sum_{i=1}^3 m_{\nu_i}^2 = 3 m_\text{lightest}^2 + 2.6 \,(4.9)\times10^{-3}~\text{eV}^2$ for normally (inversely) order neutrino masses and the lightest neutrino mass $m_\text{lightest}$. The best, model-independent bound on the absolute neutrino mass scale is set by the KATRIN experiment \cite{KATRIN:2024cdt}, $m_\beta \approx m_\text{lightest} < 0.45$~eV at 90\% confidence level. Besides requiring that $\tau_A$ exceeds the age of the universe, the DM life-time is also constrained by searches for the resulting neutrinos in oscillation experiments and neutrino telescopes, as will be discussed below.


\section{Discussion and Conclusions}
\label{sec:conclusions}

\begin{figure*}[t!]
\centering
    \includegraphics[width=0.49\textwidth]{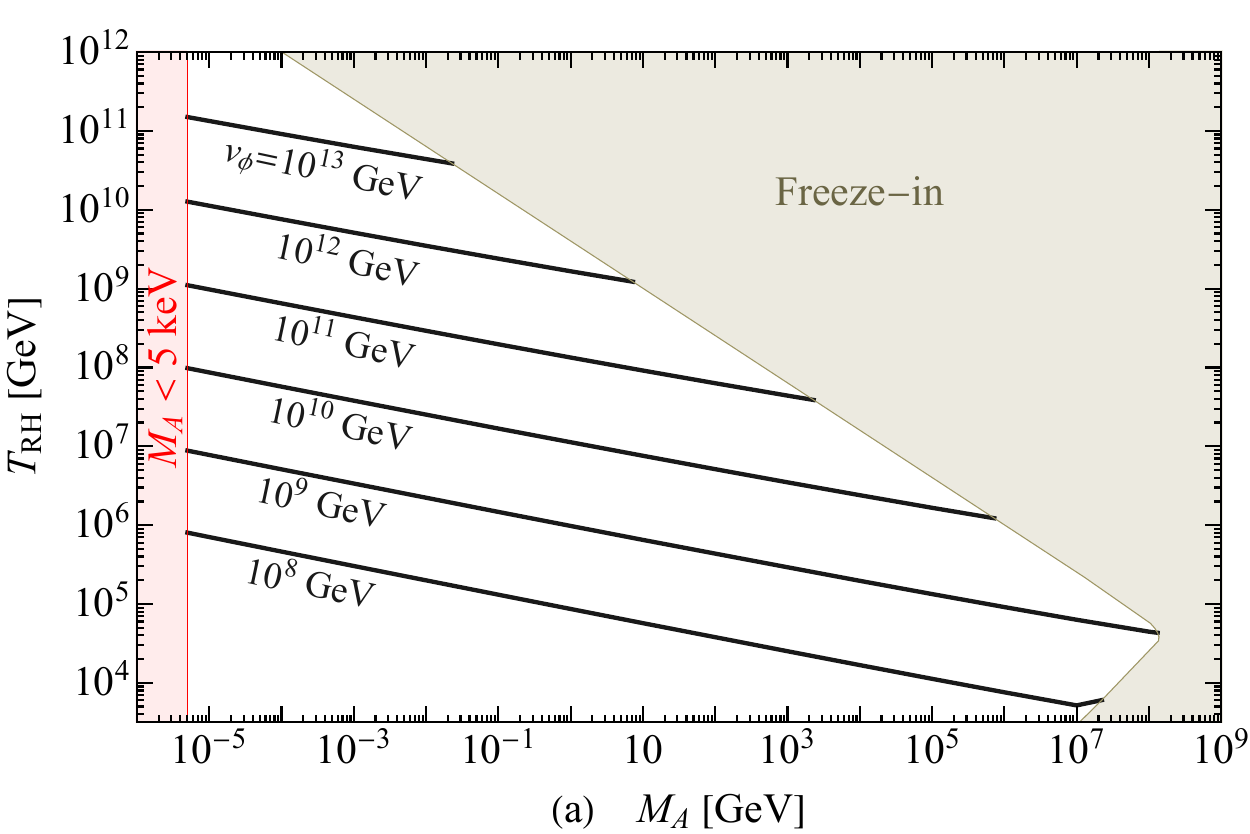}
    \includegraphics[width=0.49\textwidth]{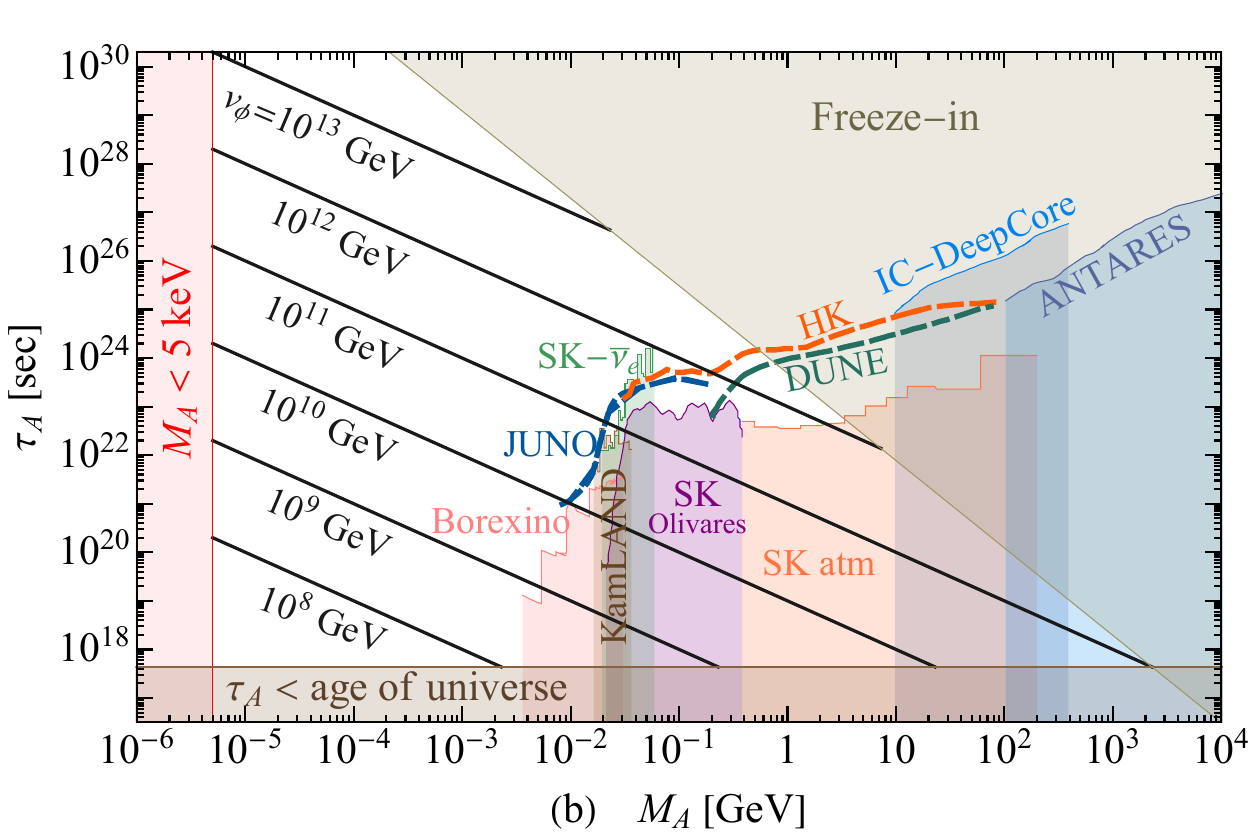}
    \caption{Reheating temperature $T_\text{RH}$ (a) and DM life-time $\tau_A$ (b) as a function of the DM mass $M_A$ for different values of $v_\phi$ (black solid lines). The other model parameters are fixed as in Eq.~\eqref{eqBM} and the observed DM relic abundance, $\Omega_Ah^2 = 0.12$, is imposed throughout. For $\tau_A$, we take the active neutrino mass as $\sum_{i=1}^3 m_{\nu_i}^2 = 10^{-2}~\text{eV}^2$. As discussed in the main text, the shaded regions are excluded from theoretical considerations and experimental constraints, whereas the dashed curves in panel~(b) indicate future sensitivities.}
\label{fig3}
\end{figure*}
Besides providing the scale of $B-L$ symmetry breaking and thus the source of light Majorana neutrino mass generation and potentially the matter-antimatter asymmetry, $v_\phi$ is also central to addressing the hierarchy problem and governing the DM relic abundance. Theoretically, within the cosmological selection mechanism, the generation of the small EW scale originates from a small mixing between $H$ and $\Phi$, naturally leading to a large hierarchy $v_\phi \gg v$, see Eq.~\eqref{eqa19}. Experimentally, the non-observation of new scalar states and significant deviations in the couplings of the Higgs from the SM values suggests that a new CP-even scalar $\phi$ must be either heavy or weakly coupled to the SM. Both possibilities typically require a large $v_\phi$, as seen in Eq.~\eqref{eq15} and Eq.~\eqref{eq16}. Future colliders such as the FCC-ee \cite{FCC:2025lpp,FCC:2025uan} are being designed to probe Higgs properties with high precision, either strengthening this conclusion or observing deviations indicating New Physics nearer the EW scale.

In Fig.~\ref{fig3}~(a), we show the reheating temperature $T_\text{RH}$ as a function of the DM mass $M_A$, for different values of the $B-L$ breaking scale $v_\phi$. We choose $\lambda_\Phi = 0.1$ as in Eq.~\eqref{eqBM} and the other parameters affect the DM phenomenology through $M_A$. The relationship arises by requiring the observed relic DM density $\Omega_Ah^2 = 0.12$~\cite{Planck:2018vyg} and there is a DM overabundance for $T_\text{RH}$ above each curve. The region with $M_A < 5$~keV is excluded by constraints from structure formation \cite{Henrich:2025gsd}.\footnote{For DM produced via UFO, this constraint can be relaxed slightly~\cite{Henrich:2025gsd} after considering the details of the model of inflation. We use the most strict limit in Fig.~\ref{fig3} for simplicity.} The shaded region in the top-right refers to the case of freeze-in DM production. While DM is feasible, this scenario depends on the initial conditions. Hence, we only focus on calculating the DM abundance for UFO. It can be seen that the predicted DM relic abundance requires a larger reheating temperature $T_\text{RH}$ for high $v_\phi$. This behavior is analogous to that observed in $Z'$-portal DM scenarios~\cite{Henrich:2025pca}. The correct DM relic abundance can thus be achieved via UFO production for high $B-L$ breaking scales and DM masses {$10~\text{MeV} \lesssim M_A \lesssim 10^8$~GeV.

As discussed, our model evades DM direct and collider searches but constraints arise from indirect searches for the DM decay $A\to\nu\nu$. The DM life-time $\tau_A = 1/\Gamma(A\to\nu\nu)$ is plotted in Fig.~\ref{fig3}~(b) as a function of $M_A$, where $\sum_{i=1}^3 m_{\nu_i}^2$ is taken as $10^{-2}~\text{eV}^2$. In addition to the constraints mentioned above, short life-times are excluded from the age of the universe $\approx 4.4\times 10^{17}$ sec~\cite{Planck:2018vyg}. The other shaded regions are excluded from existing searches or re-interpretations of searches in neutrino experiments and telescopes \cite{Arguelles:2022nbl} as indicated, namely Borexino \cite{Borexino:2019wln}, KamLAND \cite{KamLAND:2021gvi}, SuperKamiokande (SK-$\bar\nu_e$ \cite{Wan:2018ndu}, SK Olivares \cite{Olivares-DelCampo:2017feq}, SK atm \cite{Super-Kamiokande:2015qek}), Icecube (IC-DeepCore) \cite{IceCube:2021kuw} and ANTARES \cite{Albert:2016emp}. The dashed curves indicate the sensitivities of future searches \cite{Arguelles:2022nbl} at JUNO~\cite{Akita:2022lit}, DUNE~\cite{Arguelles:2019ouk} and HyperKamiokande (HK)~\cite{Bell:2020rkw}. As a result, the parameter region around $10^{10}~\text{GeV} \lesssim v_\phi \lesssim 10^{12.5}$~GeV and $10~\text{MeV} \lesssim M_A \lesssim 3$~GeV will be tested in the future.

By extending the SM with a global $U(1)_{B-L}$ symmetry, a complex scalar singlet and right-handed neutrinos, this work provides a unifying theory to address several fundamental puzzles in particle physics simultaneously: A cosmological selection mechanism to resolve the gauge hierarchy problem without introducing severe fine-tuning; the observed neutrino oscillations and baryon asymmetry can be explained via type-I seesaw and leptogenesis, respectively; and there is a viable DM candidate which is produced by ultra-relativistic freeze-out during reheating. Its relic abundance correlates directly with the reheating temperature $T_\text{RH}$. While the model evades current constraints, the predicted life-time of DM decays to active neutrinos is testable at current and next generation neutrino detectors, specifically, DUNE, HyperKamiokande and JUNO. 


\begin{acknowledgments}
The work of J.-L.~Y. has been supported by the National Natural Science Foundation of China (NNSFC) with Grants No. 12075074, No. 12235008, the Hebei Natural Science Foundation for Distinguished Young Scholars with Grant No. A2023201041 and the youth top-notch talent support program of the Hebei Province. F.~F.~D. acknowledges support from the UK Science and Technology Facilities Council (STFC) via the Consolidated Grant ST/X000613.
\end{acknowledgments}

\appendix

\section{DM Production Rate}
\label{A1}

The dominant contributions to the DM production rate $R_A(T)$ arise from the decay $\phi\to AA$ and the scattering $\phi\phi\to AA$. For the former, the DM production rate $R_A^\phi(T)$ can be written as~\cite{Giudice:2003jh}
\begin{align}
    R_A^\phi(T) &=\frac{(\lambda_\Phi v_\phi+6\kappa_3-2\kappa_4)^2}{8\pi M_\phi}K_1(M_\Phi/T),\nonumber\\
    &\approx \frac{\lambda_\Phi M_\Phi^2 T}{16\pi^3}K_1(M_\Phi/T),
\label{eqA2}
\end{align}
where $K_1(x)$ is the modified Bessel function of the second kind, and we neglect $\kappa_3$, $\kappa_4$ in the second line. 

For the scattering $\phi\phi\to AA$, we have~\cite{Giudice:2003jh}
\begin{align}
    R_A^{\phi\phi}(T) &= \frac{\lambda_\Phi^2 M_\phi T}{32\pi^5}\int_{s_{\text{min}}}^\infty \sqrt{1-4M_\phi^2/T^2}K_1\left(\frac{\sqrt s}{T}\right)\text{d}s\nonumber\\
    &\approx\frac{\lambda_\Phi^2 M_\phi T^3}{32\pi^4} e^{-2M_\phi/T}\left(1+\frac{3T}{16M_\phi}\right)
\label{eqA1}
\end{align}
where the higher order of $T/M_\phi$ are neglected approximately in the second line. Then the total DM production rate is
\begin{align}
    R_A(T) = R_A^{\phi\phi}(T) + R_A^\phi(T),
\end{align}
which is used in the Boltzmann Eq.~\eqref{eq25} in the main text to find the DM abundance. 

\bibliography{Refs}

@article{ATLAS:2012yve,
    author = "Aad, Georges and others",
    collaboration = "ATLAS",
    title = "{Observation of a new particle in the search for the Standard Model Higgs boson with the ATLAS detector at the LHC}",
    eprint = "1207.7214",
    archivePrefix = "arXiv",
    primaryClass = "hep-ex",
    reportNumber = "CERN-PH-EP-2012-218",
    doi = "10.1016/j.physletb.2012.08.020",
    journal = "Phys. Lett. B",
    volume = "716",
    pages = "1--29",
    year = "2012"
}

@article{CMS:2012qbp,
    author = "Chatrchyan, Serguei and others",
    collaboration = "CMS",
    title = "{Observation of a New Boson at a Mass of 125 GeV with the CMS Experiment at the LHC}",
    eprint = "1207.7235",
    archivePrefix = "arXiv",
    primaryClass = "hep-ex",
    reportNumber = "CMS-HIG-12-028, CERN-PH-EP-2012-220",
    doi = "10.1016/j.physletb.2012.08.021",
    journal = "Phys. Lett. B",
    volume = "716",
    pages = "30--61",
    year = "2012"
}

@article{Gildener:1976ai,
    author = "Gildener, Eldad",
    title = "{Gauge Symmetry Hierarchies}",
    reportNumber = "Print-76-0529 (HARVARD)",
    doi = "10.1103/PhysRevD.14.1667",
    journal = "Phys. Rev. D",
    volume = "14",
    pages = "1667",
    year = "1976"
}

@article{Weinberg:1978ym,
    author = "Weinberg, Steven",
    title = "{Gauge Hierarchies}",
    reportNumber = "HUTP-78/A060",
    doi = "10.1016/0370-2693(79)90248-X",
    journal = "Phys. Lett. B",
    volume = "82",
    pages = "387--391",
    year = "1979"
}

@article{Golfand:1971iw,
    author = "Golfand, Yu. A. and Likhtman, E. P.",
    editor = "Salam, A. and Sezgin, E.",
    title = "{Extension of the Algebra of Poincare Group Generators and Violation of p Invariance}",
    doi = "10.1142/9789814542340_0001",
    journal = "JETP Lett.",
    volume = "13",
    pages = "323--326",
    year = "1971"
}

@article{Ramond:1971gb,
    author = "Ramond, Pierre",
    title = "{Dual Theory for Free Fermions}",
    reportNumber = "FERMILAB-PUB-70-008-T, FERMILAB-PUB-70-008-THY, NAL-THY-8",
    doi = "10.1103/PhysRevD.3.2415",
    journal = "Phys. Rev. D",
    volume = "3",
    pages = "2415--2418",
    year = "1971"
}

@article{Wess:1973kz,
    author = "Wess, J. and Zumino, B.",
    title = "{A Lagrangian Model Invariant Under Supergauge Transformations}",
    reportNumber = "CERN-TH-1794",
    doi = "10.1016/0370-2693(74)90578-4",
    journal = "Phys. Lett. B",
    volume = "49",
    pages = "52",
    year = "1974"
}

@article{Wess:1974tw,
    author = "Wess, J. and Zumino, B.",
    editor = "Salam, A. and Sezgin, E.",
    title = "{Supergauge Transformations in Four-Dimensions}",
    doi = "10.1016/0550-3213(74)90355-1",
    journal = "Nucl. Phys. B",
    volume = "70",
    pages = "39--50",
    year = "1974"
}

@article{Arkani-Hamed:1998jmv,
    author = "Arkani-Hamed, Nima and Dimopoulos, Savas and Dvali, G. R.",
    title = "{The Hierarchy problem and new dimensions at a millimeter}",
    eprint = "hep-ph/9803315",
    archivePrefix = "arXiv",
    reportNumber = "SLAC-PUB-7769, SU-ITP-98-13",
    doi = "10.1016/S0370-2693(98)00466-3",
    journal = "Phys. Lett. B",
    volume = "429",
    pages = "263--272",
    year = "1998"
}

@article{Contino:2003ve,
    author = "Contino, Roberto and Nomura, Yasunori and Pomarol, Alex",
    title = "{Higgs as a Holographic Pseudo Goldstone Boson}",
    eprint = "hep-ph/0306259",
    archivePrefix = "arXiv",
    reportNumber = "FT-UAM-03-11, FERMILAB-PUB-03-195-T, UAB-FT-549",
    doi = "10.1016/j.nuclphysb.2003.08.027",
    journal = "Nucl. Phys. B",
    volume = "671",
    pages = "148--174",
    year = "2003"
}

@article{Agashe:2004rs,
    author = "Agashe, Kaustubh and Contino, Roberto and Pomarol, Alex",
    title = "{The Minimal composite Higgs model}",
    eprint = "hep-ph/0412089",
    archivePrefix = "arXiv",
    reportNumber = "UAB-FT-567",
    doi = "10.1016/j.nuclphysb.2005.04.035",
    journal = "Nucl. Phys. B",
    volume = "719",
    pages = "165--187",
    year = "2005"
}

@article{Agrawal:1997gf,
    author = "Agrawal, V. and Barr, Stephen M. and Donoghue, John F. and Seckel, D.",
    title = "{Viable range of the mass scale of the standard model}",
    eprint = "hep-ph/9707380",
    archivePrefix = "arXiv",
    doi = "10.1103/PhysRevD.57.5480",
    journal = "Phys. Rev. D",
    volume = "57",
    pages = "5480--5492",
    year = "1998"
}

@article{Graham:2015cka,
    author = "Graham, Peter W. and Kaplan, David E. and Rajendran, Surjeet",
    title = "{Cosmological Relaxation of the Electroweak Scale}",
    eprint = "1504.07551",
    archivePrefix = "arXiv",
    primaryClass = "hep-ph",
    doi = "10.1103/PhysRevLett.115.221801",
    journal = "Phys. Rev. Lett.",
    volume = "115",
    number = "22",
    pages = "221801",
    year = "2015"
}

@article{Csaki:2020zqz,
    author = "Cs{\'a}ki, Csaba and D'Agnolo, Raffaele Tito and Geller, Michael and Ismail, Ameen",
    title = "{Crunching Dilaton, Hidden Naturalness}",
    eprint = "2007.14396",
    archivePrefix = "arXiv",
    primaryClass = "hep-ph",
    doi = "10.1103/PhysRevLett.126.091801",
    journal = "Phys. Rev. Lett.",
    volume = "126",
    pages = "091801",
    year = "2021"
}

@article{Giudice:2019iwl,
    author = "Giudice, G. F. and Kehagias, A. and Riotto, A.",
    title = "{The Selfish Higgs}",
    eprint = "1907.05370",
    archivePrefix = "arXiv",
    primaryClass = "hep-ph",
    reportNumber = "CERN-TH-2019-114",
    doi = "10.1007/JHEP10(2019)199",
    journal = "JHEP",
    volume = "10",
    pages = "199",
    year = "2019"
}

@article{Strumia:2020bdy,
    author = "Strumia, Alessandro and Teresi, Daniele",
    title = "{Relaxing the Higgs mass and its vacuum energy by living at the top of the potential}",
    eprint = "2002.02463",
    archivePrefix = "arXiv",
    primaryClass = "hep-ph",
    doi = "10.1103/PhysRevD.101.115002",
    journal = "Phys. Rev. D",
    volume = "101",
    number = "11",
    pages = "115002",
    year = "2020"
}

@article{TitoDAgnolo:2021nhd,
    author = "Tito D'Agnolo, Raffaele and Teresi, Daniele",
    title = "{Sliding Naturalness: New Solution to the Strong-$CP$ and Electroweak-Hierarchy Problems}",
    eprint = "2106.04591",
    archivePrefix = "arXiv",
    primaryClass = "hep-ph",
    reportNumber = "CERN-TH-2021-055",
    doi = "10.1103/PhysRevLett.128.021803",
    journal = "Phys. Rev. Lett.",
    volume = "128",
    number = "2",
    pages = "021803",
    year = "2022"
}

@article{TitoDAgnolo:2021pjo,
    author = "Tito D'Agnolo, Raffaele and Teresi, Daniele",
    title = "{Sliding naturalness: cosmological selection of the weak scale}",
    eprint = "2109.13249",
    archivePrefix = "arXiv",
    primaryClass = "hep-ph",
    reportNumber = "CERN-TH-2021-138",
    doi = "10.1007/JHEP02(2022)023",
    journal = "JHEP",
    volume = "02",
    pages = "023",
    year = "2022"
}

@article{Matsedonskyi:2023tca,
    author = "Matsedonskyi, Oleksii",
    title = "{Hierarchies from landscape probability gradients and critical boundaries}",
    eprint = "2311.10139",
    archivePrefix = "arXiv",
    primaryClass = "hep-ph",
    reportNumber = "TUM-HEP 1482/23",
    doi = "10.1007/JHEP08(2024)170",
    journal = "JHEP",
    volume = "08",
    pages = "170",
    year = "2024"
}

@article{Benevedes:2025qwt,
    author = "Benevedes, Sean and Ismail, Ameen and Steingasser, Thomas",
    title = "{Gauge hierarchy and metastability from Higgs-driven crunching}",
    eprint = "2502.07876",
    archivePrefix = "arXiv",
    primaryClass = "hep-ph",
    reportNumber = "MIT-CTP/5829",
    doi = "10.1007/JHEP06(2025)228",
    journal = "JHEP",
    volume = "06",
    pages = "228",
    year = "2025"
}

@article{Geller:2018xvz,
    author = "Geller, Michael and Hochberg, Yonit and Kuflik, Eric",
    title = "{Inflating to the Weak Scale}",
    eprint = "1809.07338",
    archivePrefix = "arXiv",
    primaryClass = "hep-ph",
    doi = "10.1103/PhysRevLett.122.191802",
    journal = "Phys. Rev. Lett.",
    volume = "122",
    number = "19",
    pages = "191802",
    year = "2019"
}

@article{Chattopadhyay:2024rha,
    author = "Chattopadhyay, Susobhan and Chattopadhyay, Dibya S. and Gupta, Rick S.",
    title = "{Cosmological Selection of a Small Weak Scale from Large Vacuum Energy: A Minimal Approach}",
    eprint = "2407.15935",
    archivePrefix = "arXiv",
    primaryClass = "hep-ph",
    reportNumber = "TIFR/TH/24-14",
    doi = "10.1103/x3fg-c5p2",
    journal = "Phys. Rev. Lett.",
    volume = "134",
    number = "24",
    pages = "241803",
    year = "2025"
}

@article{Minkowski:1977sc,
    author = "Minkowski, Peter",
    title = "{$\mu \to e\gamma$ at a Rate of One Out of $10^{9}$ Muon Decays?}",
    reportNumber = "Print-77-0182 (BERN)",
    doi = "10.1016/0370-2693(77)90435-X",
    journal = "Phys. Lett. B",
    volume = "67",
    pages = "421--428",
    year = "1977"
}

@article{Weinberg:1979sa,
    author = "Weinberg, Steven",
    title = "{Baryon and Lepton Nonconserving Processes}",
    reportNumber = "HUTP-79-A050",
    doi = "10.1103/PhysRevLett.43.1566",
    journal = "Phys. Rev. Lett.",
    volume = "43",
    pages = "1566--1570",
    year = "1979"
}

@article{Fukugita:1986hr,
    author = "Fukugita, M. and Yanagida, T.",
    title = "{Baryogenesis Without Grand Unification}",
    reportNumber = "RIFP-641",
    doi = "10.1016/0370-2693(86)91126-3",
    journal = "Phys. Lett. B",
    volume = "174",
    pages = "45--47",
    year = "1986"
}

@article{Akita:2022lit,
    author = "Akita, Kensuke and Lambiase, Gaetano and Niibo, Michiru and Yamaguchi, Masahide",
    title = "{Neutrino lines from MeV dark matter annihilation and decay in JUNO}",
    eprint = "2206.06755",
    archivePrefix = "arXiv",
    primaryClass = "hep-ph",
    reportNumber = "CTPU-PTC-22-14",
    doi = "10.1088/1475-7516/2022/10/097",
    journal = "JCAP",
    volume = "10",
    pages = "097",
    year = "2022"
}

@article{Arguelles:2019ouk,
    author = {Arg{\"u}elles, Carlos A. and Diaz, Alejandro and Kheirandish, Ali and Olivares-Del-Campo, Andr{\'e}s and Safa, Ibrahim and Vincent, Aaron C.},
    title = "{Dark matter annihilation to neutrinos}",
    eprint = "1912.09486",
    archivePrefix = "arXiv",
    primaryClass = "hep-ph",
    doi = "10.1103/RevModPhys.93.035007",
    journal = "Rev. Mod. Phys.",
    volume = "93",
    number = "3",
    pages = "035007",
    year = "2021"
}

@article{Bell:2020rkw,
    author = "Bell, Nicole F. and Dolan, Matthew J. and Robles, Sandra",
    title = "{Searching for Sub-GeV Dark Matter in the Galactic Centre using Hyper-Kamiokande}",
    eprint = "2005.01950",
    archivePrefix = "arXiv",
    primaryClass = "hep-ph",
    doi = "10.1088/1475-7516/2020/09/019",
    journal = "JCAP",
    volume = "09",
    pages = "019",
    year = "2020"
}

@article{Barger:2008jx,
    author = "Barger, Vernon and Langacker, Paul and McCaskey, Mathew and Ramsey-Musolf, Michael and Shaughnessy, Gabe",
    title = "{Complex Singlet Extension of the Standard Model}",
    eprint = "0811.0393",
    archivePrefix = "arXiv",
    primaryClass = "hep-ph",
    reportNumber = "MADPH-08-1516, NUHEP-TH-08-06, ANL-HEP-PR-08-58, NPAC-08-21",
    doi = "10.1103/PhysRevD.79.015018",
    journal = "Phys. Rev. D",
    volume = "79",
    pages = "015018",
    year = "2009"
}

@article{Chen:2019ebq,
    author = "Chen, Ning and Li, Tong and Wu, Yongcheng and Bian, Ligong",
    title = "{Complementarity of the future $e^+ e^-$ colliders and gravitational waves in the probe of complex singlet extension to the standard model}",
    eprint = "1911.05579",
    archivePrefix = "arXiv",
    primaryClass = "hep-ph",
    doi = "10.1103/PhysRevD.101.075047",
    journal = "Phys. Rev. D",
    volume = "101",
    number = "7",
    pages = "075047",
    year = "2020"
}

@article{Lane:2024vur,
    author = "Lane, Samuel D. and Lewis, Ian M. and Sullivan, Matthew",
    title = "{Resonant multiscalar production in the generic complex singlet model in the multi-TeV region}",
    eprint = "2403.18003",
    archivePrefix = "arXiv",
    primaryClass = "hep-ph",
    doi = "10.1103/PhysRevD.110.055017",
    journal = "Phys. Rev. D",
    volume = "110",
    number = "5",
    pages = "055017",
    year = "2024"
}

@article{Gross:2017dan,
    author = "Gross, Christian and Lebedev, Oleg and Toma, Takashi",
    title = "{Cancellation Mechanism for Dark-Matter{\textendash}Nucleon Interaction}",
    eprint = "1708.02253",
    archivePrefix = "arXiv",
    primaryClass = "hep-ph",
    reportNumber = "HIP-2017-20-TH, TUM-HEP-1091-17, HIP-2017-20/TH, TUM-HEP/1091/17",
    doi = "10.1103/PhysRevLett.119.191801",
    journal = "Phys. Rev. Lett.",
    volume = "119",
    number = "19",
    pages = "191801",
    year = "2017"
}

@article{Chiang:2017nmu,
    author = "Chiang, Cheng-Wei and Ramsey-Musolf, Michael J. and Senaha, Eibun",
    title = "{Standard Model with a Complex Scalar Singlet: Cosmological Implications and Theoretical Considerations}",
    eprint = "1707.09960",
    archivePrefix = "arXiv",
    primaryClass = "hep-ph",
    reportNumber = "NCTS-PH-1724, ACFI-T17-16",
    doi = "10.1103/PhysRevD.97.015005",
    journal = "Phys. Rev. D",
    volume = "97",
    number = "1",
    pages = "015005",
    year = "2018"
}

@article{Cheng:2018ajh,
    author = "Cheng, Wei and Bian, Ligong",
    title = "{From inflation to cosmological electroweak phase transition with a complex scalar singlet}",
    eprint = "1801.00662",
    archivePrefix = "arXiv",
    primaryClass = "hep-ph",
    doi = "10.1103/PhysRevD.98.023524",
    journal = "Phys. Rev. D",
    volume = "98",
    number = "2",
    pages = "023524",
    year = "2018"
}

@article{Cho:2021itv,
    author = "Cho, Gi-Chol and Idegawa, Chikako and Senaha, Eibun",
    title = "{Electroweak phase transition in a complex singlet extension of the Standard Model with degenerate scalars}",
    eprint = "2105.11830",
    archivePrefix = "arXiv",
    primaryClass = "hep-ph",
    reportNumber = "OCHA-PP-366",
    doi = "10.1016/j.physletb.2021.136787",
    journal = "Phys. Lett. B",
    volume = "823",
    pages = "136787",
    year = "2021"
}

@article{Cho:2022zfg,
    author = "Cho, Gi-Chol and Idegawa, Chikako and Sugihara, Rio",
    title = "{A complex singlet extension of the standard model and multi-critical point principle}",
    eprint = "2212.13029",
    archivePrefix = "arXiv",
    primaryClass = "hep-ph",
    reportNumber = "OCHA-PP-374",
    doi = "10.1016/j.physletb.2023.137757",
    journal = "Phys. Lett. B",
    volume = "839",
    pages = "137757",
    year = "2023"
}

@article{Schicho:2022wty,
    author = "Schicho, Philipp and Tenkanen, Tuomas V. I. and White, Graham",
    title = "{Combining thermal resummation and gauge invariance for electroweak phase transition}",
    eprint = "2203.04284",
    archivePrefix = "arXiv",
    primaryClass = "hep-ph",
    reportNumber = "HIP-2022-2/TH, NORDITA 2022-009",
    doi = "10.1007/JHEP11(2022)047",
    journal = "JHEP",
    volume = "11",
    pages = "047",
    year = "2022"
}

@article{Zhang:2023mnu,
    author = "Zhang, Wenxing and Cai, Yizhou and Ramsey-Musolf, Michael J. and Zhang, Lei",
    title = "{Testing complex singlet scalar cosmology at the Large Hadron Collider}",
    eprint = "2307.01615",
    archivePrefix = "arXiv",
    primaryClass = "hep-ph",
    doi = "10.1007/JHEP01(2024)051",
    journal = "JHEP",
    volume = "01",
    pages = "051",
    year = "2024"
}

@article{Pham:2024vso,
    author = "Pham, Hieu The and Senaha, Eibun",
    title = "{Gravitational waves from domain wall collapses and dark matter in the SM with a complex scalar field}",
    eprint = "2403.16568",
    archivePrefix = "arXiv",
    primaryClass = "hep-ph",
    doi = "10.1103/PhysRevD.109.095048",
    journal = "Phys. Rev. D",
    volume = "109",
    number = "9",
    pages = "095048",
    year = "2024"
}

@article{Chikashige:1980ui,
    author = "Chikashige, Y. and Mohapatra, Rabindra N. and Peccei, R. D.",
    title = "{Are There Real Goldstone Bosons Associated with Broken Lepton Number?}",
    reportNumber = "MPI-PAE-PTH-36-80",
    doi = "10.1016/0370-2693(81)90011-3",
    journal = "Phys. Lett. B",
    volume = "98",
    pages = "265--268",
    year = "1981"
}

@article{Casas:2001sr,
    author = "Casas, J. A. and Ibarra, A.",
    title = "{Oscillating neutrinos and $\mu \to e, \gamma$}",
    eprint = "hep-ph/0103065",
    archivePrefix = "arXiv",
    reportNumber = "IEM-FT-211-01, OUTP-01-11P, IFT-UAM-CSIC-01-08",
    doi = "10.1016/S0550-3213(01)00475-8",
    journal = "Nucl. Phys. B",
    volume = "618",
    pages = "171--204",
    year = "2001"
}

@article{Ibarra:2003up,
    author = "Ibarra, A. and Ross, Graham G.",
    title = "{Neutrino phenomenology: The Case of two right-handed neutrinos}",
    eprint = "hep-ph/0312138",
    archivePrefix = "arXiv",
    reportNumber = "CERN-TH-2003-294, OUTP-0333P",
    doi = "10.1016/j.physletb.2004.04.037",
    journal = "Phys. Lett. B",
    volume = "591",
    pages = "285--296",
    year = "2004"
}

@article{Gu:2006wj,
    author = "Gu, Pei-Hong and Zhang, He and Zhou, Shun",
    title = "{A Minimal Type II Seesaw Model}",
    eprint = "hep-ph/0606302",
    archivePrefix = "arXiv",
    doi = "10.1103/PhysRevD.74.076002",
    journal = "Phys. Rev. D",
    volume = "74",
    pages = "076002",
    year = "2006"
}

@article{Hirsch:2009ra,
    author = "Hirsch, M. and Kernreiter, T. and Romao, J. C. and Villanova del Moral, Albert",
    title = "{Minimal Supersymmetric Inverse Seesaw: Neutrino masses, lepton flavour violation and LHC phenomenology}",
    eprint = "0910.2435",
    archivePrefix = "arXiv",
    primaryClass = "hep-ph",
    reportNumber = "IFIC-09-48, CFTP-09-033",
    doi = "10.1007/JHEP01(2010)103",
    journal = "JHEP",
    volume = "01",
    pages = "103",
    year = "2010"
}

@article{CarcamoHernandez:2019eme,
    author = "C{\'a}rcamo Hern{\'a}ndez, A. E. and King, S. F.",
    title = "{Littlest Inverse Seesaw Model}",
    eprint = "1903.02565",
    archivePrefix = "arXiv",
    primaryClass = "hep-ph",
    doi = "10.1016/j.nuclphysb.2020.114950",
    journal = "Nucl. Phys. B",
    volume = "953",
    pages = "114950",
    year = "2020"
}

@article{Xing:2020ald,
    author = "Xing, Zhi-zhong and Zhao, Zhen-hua",
    title = "{The minimal seesaw and leptogenesis models}",
    eprint = "2008.12090",
    archivePrefix = "arXiv",
    primaryClass = "hep-ph",
    doi = "10.1088/1361-6633/abf086",
    journal = "Rept. Prog. Phys.",
    volume = "84",
    number = "6",
    pages = "066201",
    year = "2021"
}

@article{Yang:2024duo,
    author = "Yang, Jin-Lei and Li, Jie",
    title = "{Neutrinos in the flavor-dependent U(1)F model}",
    eprint = "2411.01744",
    archivePrefix = "arXiv",
    primaryClass = "hep-ph",
    doi = "10.1103/PhysRevD.110.115007",
    journal = "Phys. Rev. D",
    volume = "110",
    number = "11",
    pages = "115007",
    year = "2024"
}

@article{Chen:2025cor,
    author = "Chen, Zi-Qiang and Hu, Xi-He and Zhou, Ye-Ling",
    title = "{Exact parametrization of a minimal seesaw model}",
    eprint = "2505.04279",
    archivePrefix = "arXiv",
    primaryClass = "hep-ph",
    doi = "10.1103/tc8x-2lp5",
    journal = "Phys. Rev. D",
    volume = "112",
    number = "11",
    pages = "115032",
    year = "2025"
}

@article{Davidson:2002qv,
    author = "Davidson, Sacha and Ibarra, Alejandro",
    title = "{A Lower bound on the right-handed neutrino mass from leptogenesis}",
    eprint = "hep-ph/0202239",
    archivePrefix = "arXiv",
    reportNumber = "OUTP-02-10P, IPPP-02-16, DCPT-02-32",
    doi = "10.1016/S0370-2693(02)01735-5",
    journal = "Phys. Lett. B",
    volume = "535",
    pages = "25--32",
    year = "2002"
}

@article{Giudice:2003jh,
    author = "Giudice, G. F. and Notari, A. and Raidal, M. and Riotto, A. and Strumia, A.",
    title = "{Towards a complete theory of thermal leptogenesis in the SM and MSSM}",
    eprint = "hep-ph/0310123",
    archivePrefix = "arXiv",
    reportNumber = "IFUP-TH-2003-37, CERN-TH-2003-240, IFUP-TH/2003-37 and CERN-TH/2003-240",
    doi = "10.1016/j.nuclphysb.2004.02.019",
    journal = "Nucl. Phys. B",
    volume = "685",
    pages = "89--149",
    year = "2004"
}

@article{Pilaftsis:2003gt,
    author = "Pilaftsis, Apostolos and Underwood, Thomas E. J.",
    title = "{Resonant leptogenesis}",
    eprint = "hep-ph/0309342",
    archivePrefix = "arXiv",
    reportNumber = "MC-TH-2003-09",
    doi = "10.1016/j.nuclphysb.2004.05.029",
    journal = "Nucl. Phys. B",
    volume = "692",
    pages = "303--345",
    year = "2004"
}

@article{Davidson:2008bu,
    author = "Davidson, Sacha and Nardi, Enrico and Nir, Yosef",
    title = "{Leptogenesis}",
    eprint = "0802.2962",
    archivePrefix = "arXiv",
    primaryClass = "hep-ph",
    doi = "10.1016/j.physrep.2008.06.002",
    journal = "Phys. Rept.",
    volume = "466",
    pages = "105--177",
    year = "2008"
}

@article{Canetti:2012kh,
    author = "Canetti, Laurent and Drewes, Marco and Frossard, Tibor and Shaposhnikov, Mikhail",
    title = "{Dark Matter, Baryogenesis and Neutrino Oscillations from Right Handed Neutrinos}",
    eprint = "1208.4607",
    archivePrefix = "arXiv",
    primaryClass = "hep-ph",
    reportNumber = "TTK-12-05, TUM-HEP-852-12, CAS-KITPC-ITP-368",
    doi = "10.1103/PhysRevD.87.093006",
    journal = "Phys. Rev. D",
    volume = "87",
    pages = "093006",
    year = "2013"
}

@article{Drewes:2016jae,
    author = "Drewes, Marco and Garbrecht, Bjorn and Gueter, Dario and Klaric, Juraj",
    title = "{Testing the low scale seesaw and leptogenesis}",
    eprint = "1609.09069",
    archivePrefix = "arXiv",
    primaryClass = "hep-ph",
    reportNumber = "TUM-HEP-1062-16",
    doi = "10.1007/JHEP08(2017)018",
    journal = "JHEP",
    volume = "08",
    pages = "018",
    year = "2017"
}

@article{Klaric:2020phc,
    author = "Klari{\'c}, Juraj and Shaposhnikov, Mikhail and Timiryasov, Inar",
    title = "{Uniting Low-Scale Leptogenesis Mechanisms}",
    eprint = "2008.13771",
    archivePrefix = "arXiv",
    primaryClass = "hep-ph",
    doi = "10.1103/PhysRevLett.127.111802",
    journal = "Phys. Rev. Lett.",
    volume = "127",
    number = "11",
    pages = "111802",
    year = "2021"
}

@article{Bhalla-Ladd:2025agq,
    author = "Bhalla-Ladd, India and Ginnett, Izzy and Tait, Tim M. P.",
    title = "{Leptogenesis during an era of early SU(2) confinement}",
    eprint = "2503.09723",
    archivePrefix = "arXiv",
    primaryClass = "hep-ph",
    doi = "10.1103/nmpt-fwz6",
    journal = "Phys. Rev. D",
    volume = "112",
    number = "7",
    pages = "075011",
    year = "2025"
}

@article{Takada:2025epa,
    author = "Takada, Rin",
    title = "{Electromagnetic leptogenesis {\textemdash} an EFT-consistent analysis via Wilson coefficients. Part I. Low-scale, non-resonant regime}",
    eprint = "2509.07698",
    archivePrefix = "arXiv",
    primaryClass = "hep-ph",
    doi = "10.1007/JHEP12(2025)010",
    journal = "JHEP",
    volume = "12",
    pages = "010",
    year = "2025"
}

@article{Okada:2025daq,
    author = "Okada, Nobuchika and Raut, Digesh",
    title = "{Analytic formulation of Leptogenesis with neutrino oscillation data employing the general parametrization for neutrino mass matrix}",
    eprint = "2506.20580",
    archivePrefix = "arXiv",
    primaryClass = "hep-ph",
    month = "6",
    year = "2025"
}

@article{Iso:2010mv,
    author = "Iso, Satoshi and Okada, Nobuchika and Orikasa, Yuta",
    title = "{Resonant Leptogenesis in the Minimal B-L Extended Standard Model at TeV}",
    eprint = "1011.4769",
    archivePrefix = "arXiv",
    primaryClass = "hep-ph",
    reportNumber = "KEK-TH-1422",
    doi = "10.1103/PhysRevD.83.093011",
    journal = "Phys. Rev. D",
    volume = "83",
    pages = "093011",
    year = "2011"
}

@article{Dev:2017wwc,
    author = "Dev, Bhupal and Garny, Mathias and Klaric, Juraj and Millington, Peter and Teresi, Daniele",
    title = "{Resonant enhancement in leptogenesis}",
    eprint = "1711.02863",
    archivePrefix = "arXiv",
    primaryClass = "hep-ph",
    reportNumber = "TUM-HEP-1110-17",
    doi = "10.1142/S0217751X18420034",
    journal = "Int. J. Mod. Phys. A",
    volume = "33",
    pages = "1842003",
    year = "2018"
}

@article{Parwani:1991gq,
    author = "Parwani, Rajesh R.",
    title = "{Resummation in a hot scalar field theory}",
    eprint = "hep-ph/9204216",
    archivePrefix = "arXiv",
    reportNumber = "ITP-SB-91-64",
    doi = "10.1103/PhysRevD.45.4695",
    journal = "Phys. Rev. D",
    volume = "45",
    pages = "4695",
    year = "1992",
    note = "[Erratum: Phys.Rev.D 48, 5965 (1993)]"
}

@article{Carrington:1991hz,
    author = "Carrington, M. E.",
    title = "{The Effective potential at finite temperature in the Standard Model}",
    reportNumber = "TPI-MINN-91-48-T-REV, TPI-MINN-91-48-T",
    doi = "10.1103/PhysRevD.45.2933",
    journal = "Phys. Rev. D",
    volume = "45",
    pages = "2933--2944",
    year = "1992"
}

@inproceedings{Quiros:1999jp,
    author = "Quiros, Mariano",
    title = "{Finite temperature field theory and phase transitions}",
    booktitle = "{ICTP Summer School in High-Energy Physics and Cosmology}",
    eprint = "hep-ph/9901312",
    archivePrefix = "arXiv",
    reportNumber = "IEM-FT-187-99",
    pages = "187--259",
    month = "1",
    year = "1999"
}

@article{Hashino:2018zsi,
    author = "Hashino, Katsuya and Kakizaki, Mitsuru and Kanemura, Shinya and Ko, Pyungwon and Matsui, Toshinori",
    title = "{Gravitational waves from first order electroweak phase transition in models with the U(1)$_{X}$ gauge symmetry}",
    eprint = "1802.02947",
    archivePrefix = "arXiv",
    primaryClass = "hep-ph",
    reportNumber = "UT-HET-123, OU-HET-957, KIAS-P18013",
    doi = "10.1007/JHEP06(2018)088",
    journal = "JHEP",
    volume = "06",
    pages = "088",
    year = "2018"
}

@article{tHooft:1979rat,
    author = "'t Hooft, Gerard",
    editor = "'t Hooft, Gerard and Itzykson, C. and Jaffe, A. and Lehmann, H. and Mitter, P. K. and Singer, I. M. and Stora, R.",
    title = "{Naturalness, chiral symmetry, and spontaneous chiral symmetry breaking}",
    reportNumber = "PRINT-80-0083 (UTRECHT)",
    doi = "10.1007/978-1-4684-7571-5_9",
    journal = "NATO Sci. Ser. B",
    volume = "59",
    pages = "135--157",
    year = "1980"
}

@article{Coleman:1973jx,
    author = "Coleman, Sidney R. and Weinberg, Erick J.",
    title = "{Radiative Corrections as the Origin of Spontaneous Symmetry Breaking}",
    doi = "10.1103/PhysRevD.7.1888",
    journal = "Phys. Rev. D",
    volume = "7",
    pages = "1888--1910",
    year = "1973"
}

@article{Binoth:1996au,
    author = "Binoth, T. and van der Bij, J. J.",
    title = "{Influence of strongly coupled, hidden scalars on Higgs signals}",
    eprint = "hep-ph/9608245",
    archivePrefix = "arXiv",
    reportNumber = "FREIBURG-THEP-96-15",
    doi = "10.1007/s002880050442",
    journal = "Z. Phys. C",
    volume = "75",
    pages = "17--25",
    year = "1997"
}

@article{Schabinger:2005ei,
    author = "Schabinger, Robert M. and Wells, James D.",
    title = "{A Minimal spontaneously broken hidden sector and its impact on Higgs boson physics at the large hadron collider}",
    eprint = "hep-ph/0509209",
    archivePrefix = "arXiv",
    doi = "10.1103/PhysRevD.72.093007",
    journal = "Phys. Rev. D",
    volume = "72",
    pages = "093007",
    year = "2005"
}

@article{Patt:2006fw,
    author = "Patt, Brian and Wilczek, Frank",
    title = "{Higgs-field portal into hidden sectors}",
    eprint = "hep-ph/0605188",
    archivePrefix = "arXiv",
    reportNumber = "MIT-CTP-3745",
    month = "5",
    year = "2006"
}

@article{Englert:2011yb,
    author = "Englert, Christoph and Plehn, Tilman and Zerwas, Dirk and Zerwas, Peter M.",
    title = "{Exploring the Higgs portal}",
    eprint = "1106.3097",
    archivePrefix = "arXiv",
    primaryClass = "hep-ph",
    doi = "10.1016/j.physletb.2011.08.002",
    journal = "Phys. Lett. B",
    volume = "703",
    pages = "298--305",
    year = "2011"
}

@article{Englert:2013gz,
    author = "Englert, Christoph and Jaeckel, Joerg and Khoze, V. V. and Spannowsky, Michael",
    title = "{Emergence of the Electroweak Scale through the Higgs Portal}",
    eprint = "1301.4224",
    archivePrefix = "arXiv",
    primaryClass = "hep-ph",
    reportNumber = "DCPT-13-04, IPPP-13-02",
    doi = "10.1007/JHEP04(2013)060",
    journal = "JHEP",
    volume = "04",
    pages = "060",
    year = "2013"
}

@article{Khoze:2013oga,
    author = "Khoze, Valentin V. and Ro, Gunnar",
    title = "{Leptogenesis and Neutrino Oscillations in the Classically Conformal Standard Model with the Higgs Portal}",
    eprint = "1307.3764",
    archivePrefix = "arXiv",
    primaryClass = "hep-ph",
    reportNumber = "DCPT-13-102, IPPP-13-51",
    doi = "10.1007/JHEP10(2013)075",
    journal = "JHEP",
    volume = "10",
    pages = "075",
    year = "2013"
}

@article{Bernal:2017kxu,
    author = "Bernal, Nicol{\'a}s and Heikinheimo, Matti and Tenkanen, Tommi and Tuominen, Kimmo and Vaskonen, Ville",
    title = "{The Dawn of FIMP Dark Matter: A Review of Models and Constraints}",
    eprint = "1706.07442",
    archivePrefix = "arXiv",
    primaryClass = "hep-ph",
    reportNumber = "PI-UAN-2017-602FT, HIP-2017-08-TH, PI-UAN--2017--602FT, HIP--2017--08-TH",
    doi = "10.1142/S0217751X1730023X",
    journal = "Int. J. Mod. Phys. A",
    volume = "32",
    number = "27",
    pages = "1730023",
    year = "2017"
}

@article{Henrich:2025sli,
    author = "Henrich, Stephen E. and Mambrini, Yann and Olive, Keith A.",
    title = "{Ultrarelativistic Freeze-Out: A Bridge from WIMPs to FIMPs}",
    eprint = "2511.02117",
    archivePrefix = "arXiv",
    primaryClass = "hep-ph",
    reportNumber = "UMN-TH-4513/25, FTPI-MINN-25/15",
    doi = "10.1103/zk9k-nbpj",
    journal = "Phys. Rev. Lett.",
    volume = "135",
    number = "22",
    pages = "221002",
    year = "2025"
}

@article{Henrich:2025gsd,
    author = "Henrich, Stephen E. and Gross, Mathieu and Mambrini, Yann and Olive, Keith A.",
    title = "{Ultrarelativistic freeze-out during reheating}",
    eprint = "2505.04703",
    archivePrefix = "arXiv",
    primaryClass = "hep-ph",
    reportNumber = "UMN-TH-4422/25, FTPI-MINN-25/04",
    doi = "10.1103/6yrm-g8t2",
    journal = "Phys. Rev. D",
    volume = "112",
    number = "10",
    pages = "103538",
    year = "2025"
}

@article{Henrich:2025pca,
    author = "Henrich, Stephen E. and Mambrini, Yann and Olive, Keith A.",
    title = "{Z' portal dark matter from post-inflationary reheating: WIMPs, FIMPs, and UFOs}",
    eprint = "2512.04229",
    archivePrefix = "arXiv",
    primaryClass = "hep-ph",
    reportNumber = "UMN-TH-4515/25, FTPI-MINN-25/16",
    doi = "10.1088/1475-7516/2026/04/068",
    journal = "JCAP",
    volume = "04",
    pages = "068",
    year = "2026"
}

@article{Cosme:2021baj,
    author = "Cosme, Catarina and Dutra, Ma{\'\i}ra and Godfrey, Stephen and Gray, Taylor R.",
    title = "{Testing freeze-in with axial and vector Z' bosons}",
    eprint = "2104.13937",
    archivePrefix = "arXiv",
    primaryClass = "hep-ph",
    doi = "10.1007/JHEP09(2021)056",
    journal = "JHEP",
    volume = "09",
    pages = "056",
    year = "2021"
}

@article{Kaneta:2019zgw,
    author = "Kaneta, Kunio and Mambrini, Yann and Olive, Keith A.",
    title = "{Radiative production of nonthermal dark matter}",
    eprint = "1901.04449",
    archivePrefix = "arXiv",
    primaryClass = "hep-ph",
    reportNumber = "LPT--Orsay 19-01, UMN--TH--3812/19, FTPI--MINN--19/03",
    doi = "10.1103/PhysRevD.99.063508",
    journal = "Phys. Rev. D",
    volume = "99",
    number = "6",
    pages = "063508",
    year = "2019"
}

@article{Hannestad:2004px,
    author = "Hannestad, Steen",
    title = "{What is the lowest possible reheating temperature?}",
    eprint = "astro-ph/0403291",
    archivePrefix = "arXiv",
    doi = "10.1103/PhysRevD.70.043506",
    journal = "Phys. Rev. D",
    volume = "70",
    pages = "043506",
    year = "2004"
}

@article{Mohapatra:2021ozu,
    author = "Mohapatra, Rabindra N. and Okada, Nobuchika",
    title = "{Unified model for inflation, pseudo-Goldstone dark matter, neutrino mass, and baryogenesis}",
    eprint = "2112.02069",
    archivePrefix = "arXiv",
    primaryClass = "hep-ph",
    doi = "10.1103/PhysRevD.105.035024",
    journal = "Phys. Rev. D",
    volume = "105",
    number = "3",
    pages = "035024",
    year = "2022"
}

@article{Esteban:2024eli,
    author = "Esteban, Ivan and Gonzalez-Garcia, M. C. and Maltoni, Michele and Martinez-Soler, Ivan and Pinheiro, Jo{\~a}o Paulo and Schwetz, Thomas",
    title = "{NuFit-6.0: updated global analysis of three-flavor neutrino oscillations}",
    eprint = "2410.05380",
    archivePrefix = "arXiv",
    primaryClass = "hep-ph",
    reportNumber = "IFT-UAM/CSIC-24-140, YITP-SB-2024-24, IPPP/24/64, IPPP/24/64, IFT-UAM/CSIC-24-140, YITP-SB-2024-24",
    doi = "10.1007/JHEP12(2024)216",
    journal = "JHEP",
    volume = "12",
    pages = "216",
    year = "2024"
}

\end{document}